\documentclass[sigconf]{acmart}
\usepackage[utf8]{inputenc}
\usepackage{adjustbox}
\usepackage[ruled,vlined]{algorithm2e}
\usepackage{algorithmic}
\usepackage{graphicx}
\usepackage{textcomp}
\usepackage{xcolor}
\usepackage{balance}
\usepackage{enumitem}
\usepackage{setspace}
\usepackage{wrapfig}
\usepackage{listings}
\usepackage{color}
\usepackage{caption}
\usepackage{subcaption}
\usepackage{tikz}
\usetikzlibrary{shapes.geometric, arrows}
\usetikzlibrary{fit}
\usetikzlibrary{calc}
\tikzstyle{process_g} = [rectangle, minimum width=2.3cm, minimum height=0.75cm, text centered, text width=2.3cm, draw=black, fill=gray!40]

\tikzstyle{process} = [rectangle, minimum width=2.3cm, minimum height=0.75cm, text centered, text width=2.3cm, draw=black, fill=gray!10]
\tikzstyle{arrow} = [thick,->,>=stealth]

\newcommand{\etal}{\emph{et al. }}
\newcommand{\ie}{\emph{i.e.}, }
\newcommand{\eg}{\emph{e.g.}, }

\newcommand{\cf}{\emph{cf.}\xspace}
\newcommand{\HAS}{\emph{HTTP Adaptive Streaming }}

\newcommand{\DCT}{\emph{Discrete Cosine Transform }}

\newcommand{\VQA}{\emph{Video Quality Assessment}}
\newcommand{\JND}{\emph{Just Noticeable Difference}}

\newcommand{\tvpm}{\texttt{TQPM}\xspace}
\newcommand{\rf}[1]{\textcolor{black}{#1}}

\newcommand{\sota}{state-of-the-art\xspace}

\copyrightyear{2023}
\acmYear{2023}
\setcopyright{rightsretained}
\acmConference[MHV '23]{Mile-High Video Conference}{May 7--10, 2023}{Denver, CO, USA}
\acmBooktitle{Mile-High Video Conference (MHV '23), May 7--10, 2023, Denver, CO, USA}
\acmDOI{10.1145/3588444.3591012}
\acmISBN{979-8-4007-0160-3/23/05}

\begin{document}

\title{Transcoding Quality Prediction for Adaptive Video Streaming}

\author{Vignesh V Menon}
\email{vignesh.menon@aau.at}
\orcid{0000-0003-1454-6146}
\affiliation{
  \institution{\small{Christian Doppler Laboratory ATHENA}}
  \institution{Institute of Information Technology (ITEC)}
  \institution{Alpen-Adria-Universität Klagenfurt}
  \city{Klagenfurt}
  \country{Austria}
}

\author{Reza Farahani}
\email{reza.farahani@aau.at}
\orcid{0000-0002-2376-5802}
\affiliation{
  \institution{\small{Christian Doppler Laboratory ATHENA}}
  \institution{Institute of Information Technology (ITEC)}
  \institution{Alpen-Adria-Universität Klagenfurt}
  \city{Klagenfurt}
  \country{Austria}
}

\author{Prajit T Rajendran}
\email{prajit.thazhurazhikath@cea.fr}
\orcid{0000-0002-8283-9891}
\affiliation{
  \institution{\small{CEA, List, F-91120 Palaiseau}}
  \institution{Institute of Information Technology (ITEC)}
  \institution{Université Paris-Saclay}
  \city{Paris}
  \country{France}
}

\author{Mohammad Ghanbari}
\email{ghan@essex.ac.uk}
\orcid{0000-0002-5482-8378}
\affiliation{
  \institution{\small{School of Computer Science and Electronic Engineering}}
  \institution{University of Essex}
  \city{Colchester}
  \country{UK}
}

\author{Hermann Hellwagner}
\email{hermann.hellwagner@aau.at}
\orcid{0000-0003-1114-2584}
\affiliation{
  \institution{\small{Christian Doppler Laboratory ATHENA}}
  \institution{Institute of Information Technology (ITEC)}
  \institution{Alpen-Adria-Universität Klagenfurt}
  \city{Klagenfurt}
  \country{Austria}
}

\author{Christian Timmerer}
\email{christian.timmerer@aau.at}
\orcid{0000-0002-0031-5243}
\affiliation{
  \institution{\small{Christian Doppler Laboratory ATHENA}}
  \institution{Institute of Information Technology (ITEC)}
  \institution{Alpen-Adria-Universität Klagenfurt}
  \city{Klagenfurt}
  \country{Austria}
}


\renewcommand{\shortauthors}{Vignesh V Menon~\etal}

\begin{abstract}
In recent years, video streaming applications have proliferated the demand for \VQA~(VQA). \textit{Reduced reference} video quality assessment (RR-VQA) is a category of VQA where certain features (\eg texture, edges) of the original video are provided for quality assessment. It is a popular research area for various applications such as social media, online games, and video streaming. This paper \rf{introduces} a \rf{\textit{reduced reference}} \textbf{T}ranscoding \textbf{Q}uality \textbf{P}rediction \textbf{M}odel (\tvpm) to determine the visual quality score of the video possibly transcoded in multiple stages. The quality is predicted using \DCT (DCT)-energy-based features of the video (\ie the video's brightness, spatial texture information, and temporal activity) and the target bitrate representation of each transcoding stage. \rf{To do that, the problem is formulated, and a \textit{Long Short-Term Memory} (LSTM)-based quality prediction model is presented.} Experimental results \rf{illustrate} that, on average, \tvpm yields PSNR, SSIM, and VMAF predictions with an $R^2$ score of 0.83, 0.85, and 0.87, respectively, and \rf{\textit{Mean Absolute Error}} (MAE) of 1.31 dB, 1.19 dB, and 3.01, respectively, for single-stage transcoding. Furthermore, an $R^2$ score of 0.84, 0.86, and 0.91, respectively, and MAE of 1.32 dB, 1.33 dB, and 3.25, respectively, are observed for a two-stage transcoding scenario. Moreover, the average processing time of \tvpm for 4s segments is 0.328s, making it a practical VQA method in online streaming applications. 
\end{abstract}



\maketitle

\section{Introduction}
\begin{figure}[t]
\centering
\includegraphics[width=0.405\textwidth]{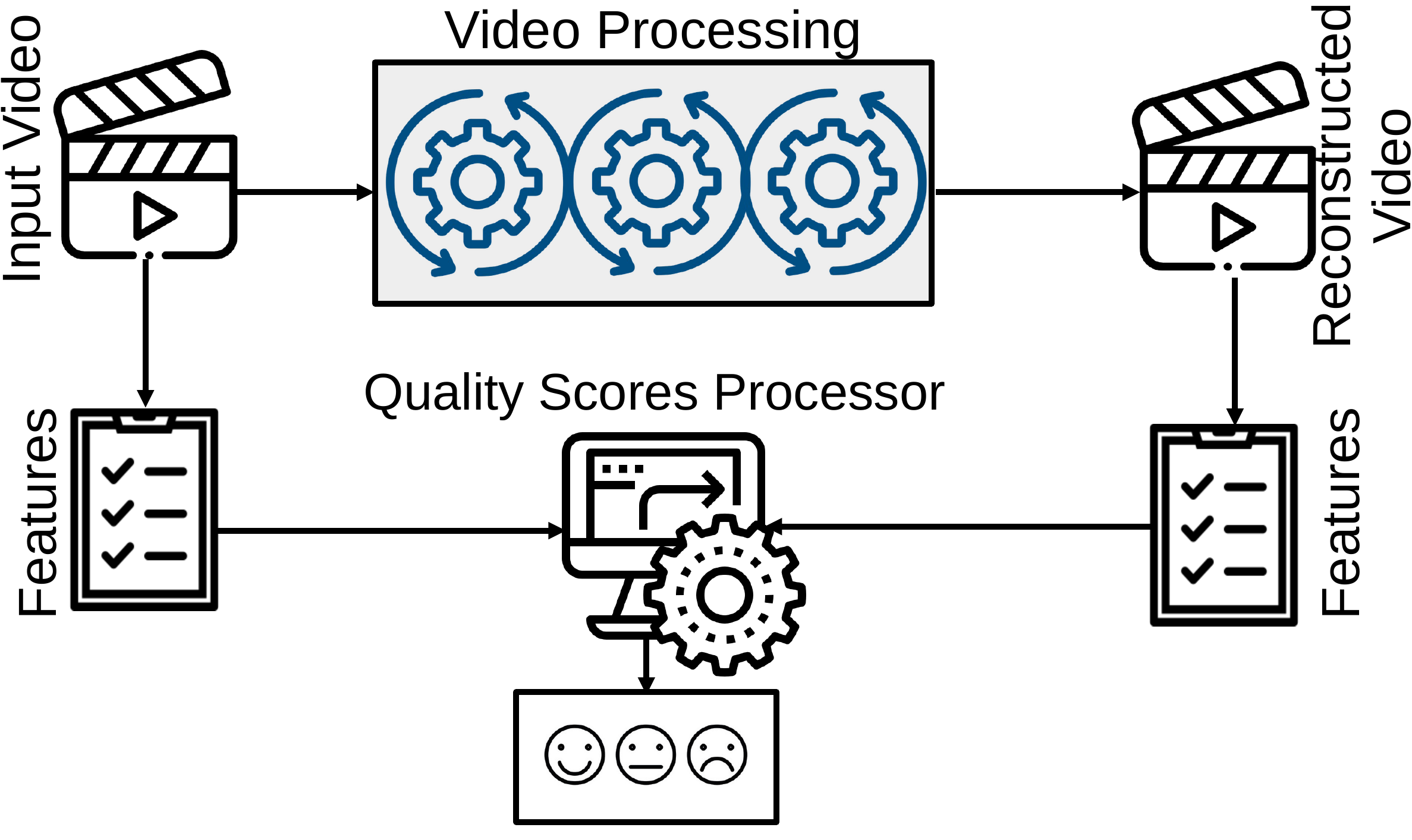}
\caption{Workflow of state-of-the-art RR-VQA methods.}
\label{fig:rrvqa_sota}
\end{figure}
The demand for \VQA~(VQA) is growing in video streaming applications. It plays an essential role in video processing \rf{from capturing to rendering, including} compression, transmission, restoration, and display~\cite{VQ_survey}. With all the available encoding options and trade-offs to consider in \HAS~(HAS)~\cite{DASH_Survey}, \rf{having} a \rf{lightweight}, and reliable VQA method is crucial. 
According to the degree of information available for the reference video signals, VQA is classified into \textit{full reference} (FR), \textit{reduced reference} (RR), and \textit{no reference} (NR) methods. NR-VQA methods are ``blind'', where the original video content is not used for quality assessment, leading to an unreliable VQA~\cite{VQ_survey}. On the other hand, since RR-VQA methods use \textit{(i)} less overhead data compared to FR-based VQA approaches and  \textit{(ii)} are more reliable than NR-based VQA methods, they are employed in real-time scenarios~\cite{rr_vqa_survey}.

The workflow of the \rf{\sota} RR-VQA methods is shown in Figure~\ref{fig:rrvqa_sota}. The characteristic features of the original video and the reconstructed video (\eg pixels, relative entropy or entropy difference~\cite{rr_vqa1,rr_vqa2}, frequency domain features like DCT~\cite{rr_dct1}) after any arbitrary video processing process are extracted. The quality score processor (mostly ML-based implementations in the literature) combines these features to predict the resultant video quality~\cite{rr_vqa_survey}.
Since \textit{Peak Signal to Noise Ratio} (PSNR) remains the \textit{de facto} industry standard for video quality evaluation, many RR-VQA methods are developed to evaluate it~\cite{rr_psnr1,rr_psnr2}. Furthermore, there are methods that predict the \rf{Structural Similarity Index (SSIM)}~\cite{rr_ssim1, rr_ssim2}, Spatio-temporal RR Entropic Differences (STRRED)~\cite{rr_strred1}, and Spatial RR Entropic Differences (SRRED)~\cite{rr_srred1} metrics. However, the metrics mentioned above have limitations, such as neglecting the temporal nature of compression artifacts~\cite{vmaf_ref1}. 
To bridge these gaps, \textit{Video Multi-method Assessment Fusion} (VMAF) was introduced~\cite{VMAF}. VMAF was proposed as an FR-VQA model that combines quality-aware features to predict perceptual quality. \rf{For that, it incorporates} human vision modeling with machine learning and \rf{offers an acceptable} prediction of the video QoE~\cite{vmaf_ref1}. VMAF is an optimization criterion for better encoding decisions in different applications. \rf{As an} example, Orduna~\etal~\cite{vmaf_ref2} prove that VMAF can be used without any specific training or adjustments to obtain the quality of 360-degree virtual reality (VR) sequences perceived by users. Zadtootaghaj~\etal \cite{vmaf_ref3} use VMAF to analyze the video quality of \rf{online video} gaming services and calculate the minimum encoding bitrate to \rf{reduce the required bandwidth of different streaming games significantly.} Sakaushi~\etal \cite{vmaf_ref4} present a video surveillance system where VMAF is used to measure how the quality of the video is degraded for different bitrates. In ~\cite{opte_ref}, optimized bitrate-resolution pairs that maximize VMAF are selected for the bitrate ladder. In ~\cite{ppte_ref, jtps_ref}, perceptually-aware optimized bitrate-resolution pairs that maximize the visual quality and compression efficiency are selected for the bitrate ladder. Additionally, in ~\cite{coda_ref}, the optimized framerate that yields the highest VMAF is selected for every target bitrate in the ladder. Hence, visual quality prediction enables the server to choose the optimized encoding parameters for the bitrate ladder~\cite{vqa_icip_ref}.  

\textbf{\textit{Contributions:}} This paper proposes a reduced-reference transcoding quality prediction model (\tvpm) for video streaming applications. To the best of our knowledge, this is the first work proposed to \textit{predict VMAF for multi-stage transcoding}, especially in video streaming applications, where the video segment is subjected to multiple stages of transcoding before being transcoded to the target bitrate representation. To do that, first, DCT-energy-based features are extracted from the input video segment, and the information of the transcoding pipeline (\ie target bitrate representation of encoder in each stage) is used as the \textit{reduced reference} for VMAF prediction. \text{Next}, feature extraction is carried out only for the input video segment. This method contrasts the \rf{\sota} RR-VQA methods where feature extraction is carried out for the input and the output video segments from the transcoding system. The prediction performance of the proposed model is validated using \rf{\textit{Apple HTTP Live Streaming} (HLS)} bitrate ladder\footnote{\label{apple_hls_ref}\href{https://developer.apple.com/documentation/http\_live\_streaming/hls\_authoring\_specification\_for\_apple\_devices}{https://developer.apple.com/documentation/http\_live\_streaming/ hls\_authoring\_specification\_for\_apple\_devices}, last access: Apr 02, 2023.} transcoding using the x265\footnote{\label{x265_ref}\href{https://www.videolan.org/developers/x265.html}{https://www.videolan.org/developers/x265.html}, last access: Apr 02, 2023.} HEVC~\cite{HEVC} open source encoder.

\begin{figure}[t]
\centering
\includegraphics[width=.47\textwidth]{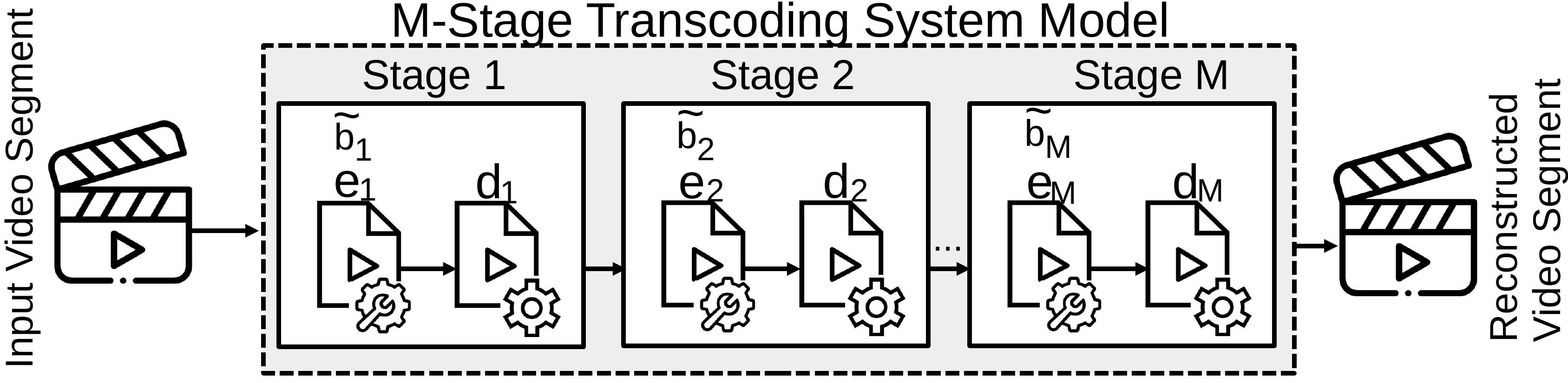}
\caption{\texttt{M}-stage transcoding model considered in this paper. Here, $e_{i}$ and $d_{i}$ represent the encoding and decoding in $i^{th}$ stage of transcoding, while $\Tilde{b_{i}}$ denotes the target bitrate of $e_{i}$ where $i \in [1, M]$.}
\label{fig:mtrans_model}
\end{figure}

\textbf{\textit{Paper outline:}} Section~\ref{sec:mstage_model} explains the M-stage transcoding model formulated in this paper, while Section~\ref{sec:prop_arch} discusses the architecture of \tvpm. Section~\ref{sec:evaluation} illustrates the evaluation of the \tvpm performance. Finally, Section~\ref{sec:conclusion_future_dir} concludes the paper.

\begin{figure*}[t]
\centering
\includegraphics[width=.785\textwidth]{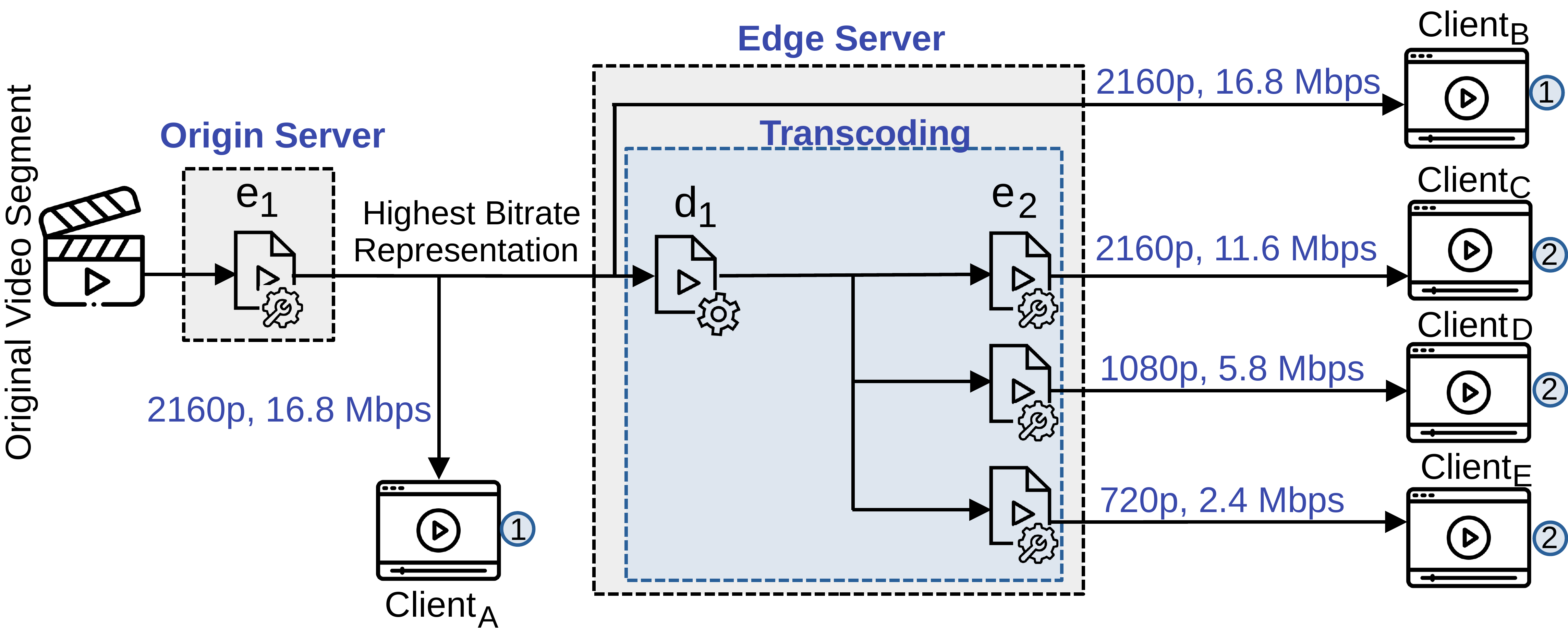}
\caption{An example scenario of VQA in adaptive streaming applications. Clients A and B receive the highest bitrate representation of the bitrate ladder, encoded at the origin server (single-stage transcoding), while Clients C, D, and E receive lower bitrate representations transcoded at the edge server (two-stage transcoding).}
\label{fig:motivation_m_stage}
\end{figure*}

\section{M-stage transcoding model}
\label{sec:mstage_model}
Recently, video transcoding has been considered a prevalent solution for reconstructing video sequences at \textit{in-network servers} (deployed at cloud or edge) in latency-sensitive video streaming applications~\cite{farahani2021cdn,farahani2022richter,farahani2022leader,farahani2022ararat}. Hence, in this paper, a generalized \texttt{M}-stage transcoding model for HAS as depicted in Figure~\ref{fig:mtrans_model} is proposed, targeting the following scenarios:
\begin{enumerate}
    \item \textit{Single-stage transcoding:} This is the scenario where the user receives the bitstream produced by the source server. As shown in Figure~\ref{fig:motivation_m_stage}, clients A and B receive the bitrate representation generated at the origin server. Here, VQA can be accomplished at the origin server, as the original and reconstructed video segments are available at the origin server. However, in the state-of-the-art VQA methods, the encoding process must be complete to determine the visual quality score. Moreover, the time taken for feature extraction ($\tau_{f}$) of the input and reconstructed video segments adds to the latency. 
    
    \item \textit{Two-stage transcoding:} In these applications, a higher bitrate representation already available in the edge server is transcoded to a lower bitrate representation requested by the user. As shown in Figure~\ref{fig:motivation_m_stage}, clients C, D, and E receive 11.6 Mbps, 5.8 Mbps, and 2.4 Mbps representations. The edge server transcodes the video segment from the 16.8 Mbps representations to the requested representations. In this manner, the response delay and the backhaul traffic between the origin and the edge servers is expected to be reduced~\cite{farahani2021cdn}. State-of-the-art VQA methods cannot be used in this scenario as \textit{(i)} the original input video segment is not available as the reference at the destination (client) and \textit{(ii)} the final reconstructed video segment is not available at the source (origin server). Assuming a hypothetical scenario where the original and reconstructed video segments are available together at the source or destination, the total processing time would include two encoding and decoding steps and feature extraction of the original and reconstructed segments.
\end{enumerate}

There shall be scenarios of three-stage transcoding that involve two edge servers. As depicted in Figure~\ref{fig:mtrans_model}, the generalized \texttt{M}-stage transcoding model for HAS consists of a series of \texttt{M} encoders and \texttt{M} decoders in a chain. \texttt{M=1} transcoding corresponds to the single-stage transcoding while \texttt{M=2} transcoding corresponds to the two-stage transcoding.
As explained, RR-VQA poses numerous problems while deployed in multi-stage transcoding applications. First, the total transcoding latency to compute video quality ($\tau_{T}$) using the input and the final reconstructed video segments is very high. This is because of the encoding and decoding times in the \texttt{M}-stage transcoding process (\texttt{M} encoding and \texttt{M} decoding processes), \rf{plus the} time taken for feature extraction ($\tau_{f}$) of the input and reconstructed video segments add to the latency. \rf{The total transcoding latency is formulated in Eq.~(\ref{eq:tot_ref_lat}), where $\tau_{e_{i}}$ and $\tau_{d_{i}}$ represents the time taken to encode and decode at the $i^{th}$ transcoding stage, respectively.}
\begin{equation}
\label{eq:tot_ref_lat}
\tau_{T} = \sum_{i=1}^{M}(\tau_{e_{i}} + \tau_{d_{i}}) + 2\cdot \tau_{f}
\end{equation}

Second, determining VMAF is cumbersome in most video streaming applications where \rf{\textit{(i)}} the original input video segment is not available as the reference at the destination; \rf{\textit{(ii)}} the final reconstructed video segment is not available at the source; \rf{\textit{(iii)} slow VMAF decision-making is not acceptable for online latency-sensitive services.} VQA at source by predicting VMAF using the input video segment characteristics and the transcoding system characteristics solves the abovementioned problems.

\section{\tvpm Architecture}
\label{sec:prop_arch}
\begin{figure}[t]
\centering
\includegraphics[width=0.42\textwidth]{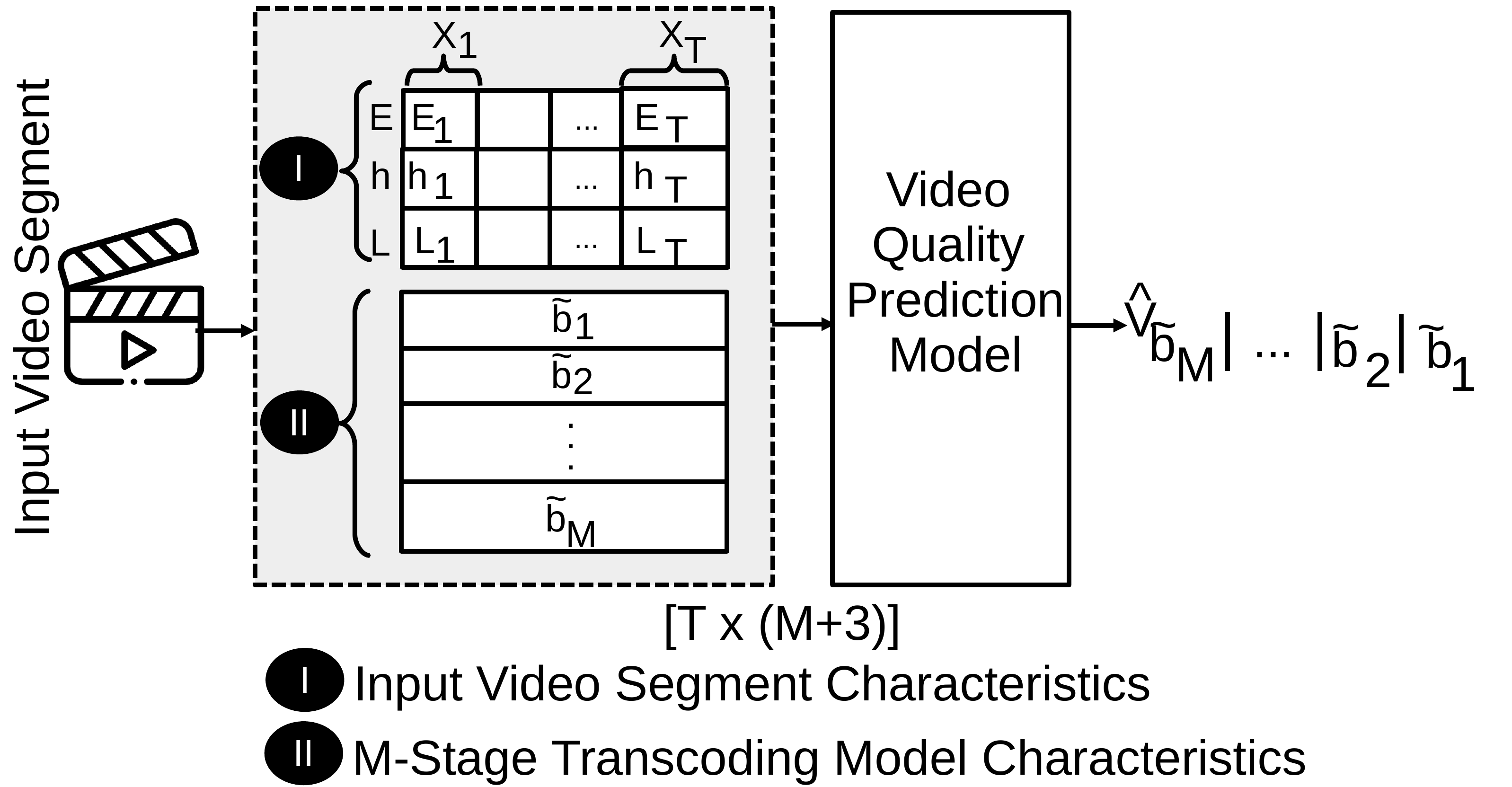}
\vspace{0pt}
\caption{\tvpm architecture}
\label{fig:tqpm_arch}
\vspace{-1.5em}
\end{figure}

\rf{The \tvpm architecture is shown in Figure~\ref{fig:tqpm_arch}, which comprises} three steps:
\begin{enumerate}[leftmargin=*]
  \item input video segment characterization (Section~\ref{sec:feature_extraction})
  \item transcoding model Characterization (Section~\ref{sec:vmaf_pred})
  \item video quality prediction (Section~\ref{sec:vmaf_pred})
\end{enumerate}

Selecting low-complexity features to characterize the input video segment is critical to utilize lightweight prediction models for quality prediction. High-complexity features would require heavier models (in terms of model size and inference time), contributing to prediction latency. \rf{Extracting \sota} \rf{\textit{Spatial Information}} (SI) and \rf{\textit{Temporal Information}} (TI) features are computationally intensive \rf{tasks} and do not correlate well with the transcoded video quality~\cite{vca_ref}. This paper uses a \rf{lightweight} and low-latency feature extraction from input video segments as explained in Section~\ref{sec:feature_extraction}. The extracted features, along with the encoding target bitrate representations for each stage of the transcoding process, \ie $\Tilde{b}_{1}$, $\Tilde{b}_{2}$,.., $\Tilde{b}_{M}$, are \rf{employed} to predict the visual quality, in terms of PSNR, SSIM, and VMAF, as discussed in Section~\ref{sec:vmaf_pred}.

\subsection{Input Video Segment Characterization}
\label{sec:feature_extraction}
Three DCT-energy-based features extracted by the \textit{Video Complexity Analyzer} (VCA)~\cite{vca_ref} open-source software, \ie \textit{(i)} the average texture energy $E$, \textit{(ii)} the average temporal energy $h$, and \textit{(iii)} the average luminescence $L$ are used as the \textit{reduced reference} for each video segment. These features are based on the luma channel of the video segment. Chroma channels are not considered in the proposed solution since the rate control of most of the \sota encoders does not consider them. Furthermore, VQA metrics like VMAF emphasize the luma channel more than the chroma channels. The features are based on our previous work~\cite{vca_ref} and are included here to have the paper self-contained.   
Firstly, the texture of every non-overlapping block $k$ in each frame $p$ is calculated \rf{using Eq.~(\ref{eq:block_energy})}:
\begin{equation}
\label{eq:block_energy}
H_{p, k} = \sum_{i=0}^{w-1} \sum_{j=0}^{w-1} e^{|(\frac{ij}{w^{2}})^2 -1|}|D(i,j)|
\end{equation}
where $w \times w$ pixels is the size of the block, and $D(i,j)$ is the $(i, j)^{th}$ DCT component when $i + j > 0$, and 0 otherwise~\cite{dct_ref}. The texture is averaged to determine the \textit{spatial energy} feature per frame, \rf{\ie $E_{p}$}, as shown \rf{in Eq.~(\ref{eq:spatial_energy})}:
\begin{equation}
\label{eq:spatial_energy}
E_{p} = \sum_{k=0}^{K-1} \frac{H_{p, k}}{K \cdot w^{2}}
\end{equation}
where $K$ represents the number of blocks in the frame $p$~\cite{mmsp_paper_ref}. Furthermore, the block-wise sum of absolute difference (SAD) of the texture energy of each frame compared to its previous frame is computed and then averaged per frame to obtain the \textit{temporal energy} feature per frame, (\ie $h_p$) illustrated in Eq.~(\ref{eq:4}):
\begin{equation}
h_{p} = \sum_{k=0}^{K-1} \frac{\mid H_{p, k} - H_{p-1, k}\mid}{K \cdot w^{2}}
\label{eq:4}
\end{equation}

The luminescence of non-overlapping blocks $k$ of each frame $p$ is defined as:
\begin{equation}
L_{p, k} = \sqrt{DCT(0,0)}
\end{equation}
where $DCT(0,0)$ is the $DC$ component in the DCT calculation. \rf{Moreover}, the block-wise luminescence is averaged per frame denoted as $L_{p}$ as shown \rf{in Eq.~(\ref{eq:luma})}.
\begin{equation}
\label{eq:luma}
L_{p} = \sum_{k=0}^{K-1} \frac{L_{p, k}}{K \cdot w^{2}}
\end{equation}

The video segment is divided into $T$ chunks with a fixed number of frames \rf{(\ie $f_c$)} in each chunk. The averages of the $E$, $h$, and $L$ features of each chunk are computed to obtain the \textit{reduced reference representation} of the input video segment, expressed as:
\begin{equation}
X = \{x_{1}, x_{2},.., x_{T}\}
\end{equation}
where, $x_{i}$ is the feature set of every $i^{th}$ chunk, represented as :
\begin{equation}
x_{i} = [E_{i}, h_{i}, L_{i}] \hspace{0.4cm} \forall i \in [1,T]
\end{equation}

\subsection{Video Quality Prediction}
\label{sec:vmaf_pred}
For the sake of simplicity, the settings of the encoders in the \texttt{M}-stage transcoding process, except the target bitrate-resolution pair, are assumed identical~\cite{emes_ref}. The resolutions corresponding to the target bitrates in the bitrate ladder are also assumed to be fixed. Therefore, the transcoding model can be characterized as follows:
\begin{equation}
\Tilde{B} = [\Tilde{b}_{1}, \Tilde{b}_{2},.., \Tilde{b}_{M}]
\end{equation}
where $\Tilde{b}_{i}$ represents the target bitrate of the $e_{i}$ encoder (\cf Fig.~\ref{fig:mtrans_model}). Note that $\Tilde{B}$ is appended to $x_{i}$, which is determined during the input video segment characterization phase, to obtain:

\begin{equation}
\Tilde{x_{i}} = [x_{i} | \Tilde{B}]^{T} \hspace{0.4cm} \forall \Tilde{x_{i}} \in \Tilde{X}, \hspace{0.4cm} i \in [1, T]
\end{equation}
The predicted quality $\hat{v}_{\Tilde{b}_{M}|..|\Tilde{b}_{1}}$ can be presented as:
\begin{equation}
\hat{v}_{\Tilde{b}_{M}|..|\Tilde{b}_{1}} = f(\Tilde{X})
\end{equation}
LSTM models are typically used in time series prediction applications and can mitigate essential issues in long-term prediction, such as vanishing or exploding gradients~\cite{lstm_survey_ref}. Thus, an LSTM-based prediction model~\cite{lstm_ref} is used in this work. The described features are input to the model~\cite{lstm_ref} as a vector of dimension $[T\times(M+3)]$, where $T$ denotes the number of chunks in the video segment. More specifically, the feature sequences in the series $\Tilde{X}$ are input to the LSTM model, which predicts visual quality for the corresponding input video segment and chain of encoders in the transcoding process.

The upper bound for the acceptable deviation from the ground truth quality is considered to be one \JND~(JND),
\begin{equation}
|\hat{v} - v_{G}| < 1 \text{JND}
\end{equation}
where $\hat{v}$ and $v_{G}$ are the predicted and the ground truth quality, respectively. In this paper, the average target JND is considered as six VMAF points\footnote{\href{https://streaminglearningcenter.com/codecs/finding-the-just-noticeable-difference-with-netflix-vmaf.html}{https://streaminglearningcenter.com/codecs/finding-the-just-noticeable-difference-with-netflix-vmaf.html}, last access: Apr 02, 2023.} based on current industry practices.

\section{Evaluation}
\label{sec:evaluation}
This section first explains the evaluation setup and then presents the experimental results.
\begin{figure*}[t]
\centering
\begin{subfigure}{0.24\textwidth}
    \centering
    \includegraphics[width=\textwidth]{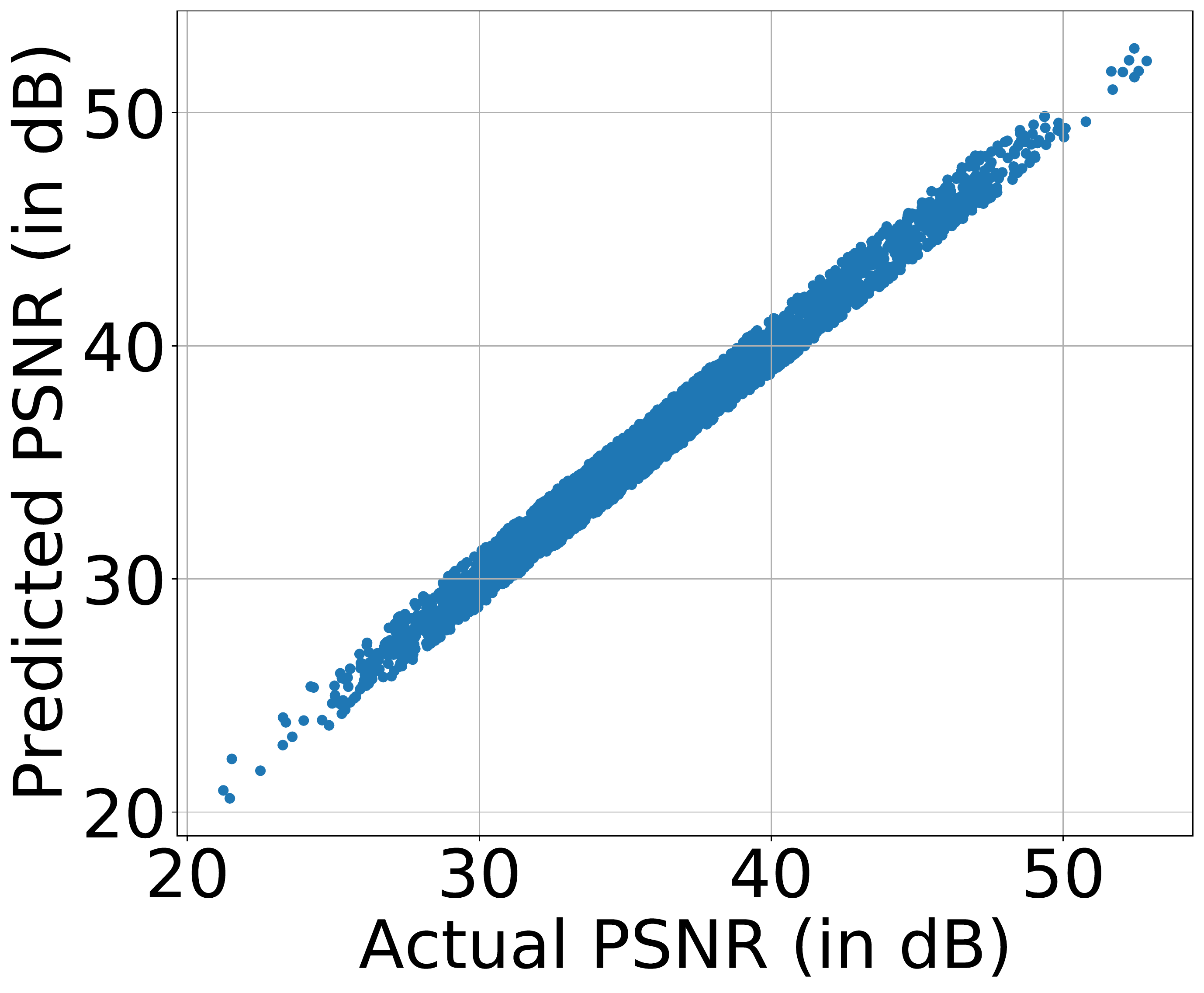}
    \caption{}
    \label{fig:psnr_m1_scatter}    
\end{subfigure}
\hfill
\begin{subfigure}{0.24\textwidth}
    \centering
    \includegraphics[width=\textwidth]{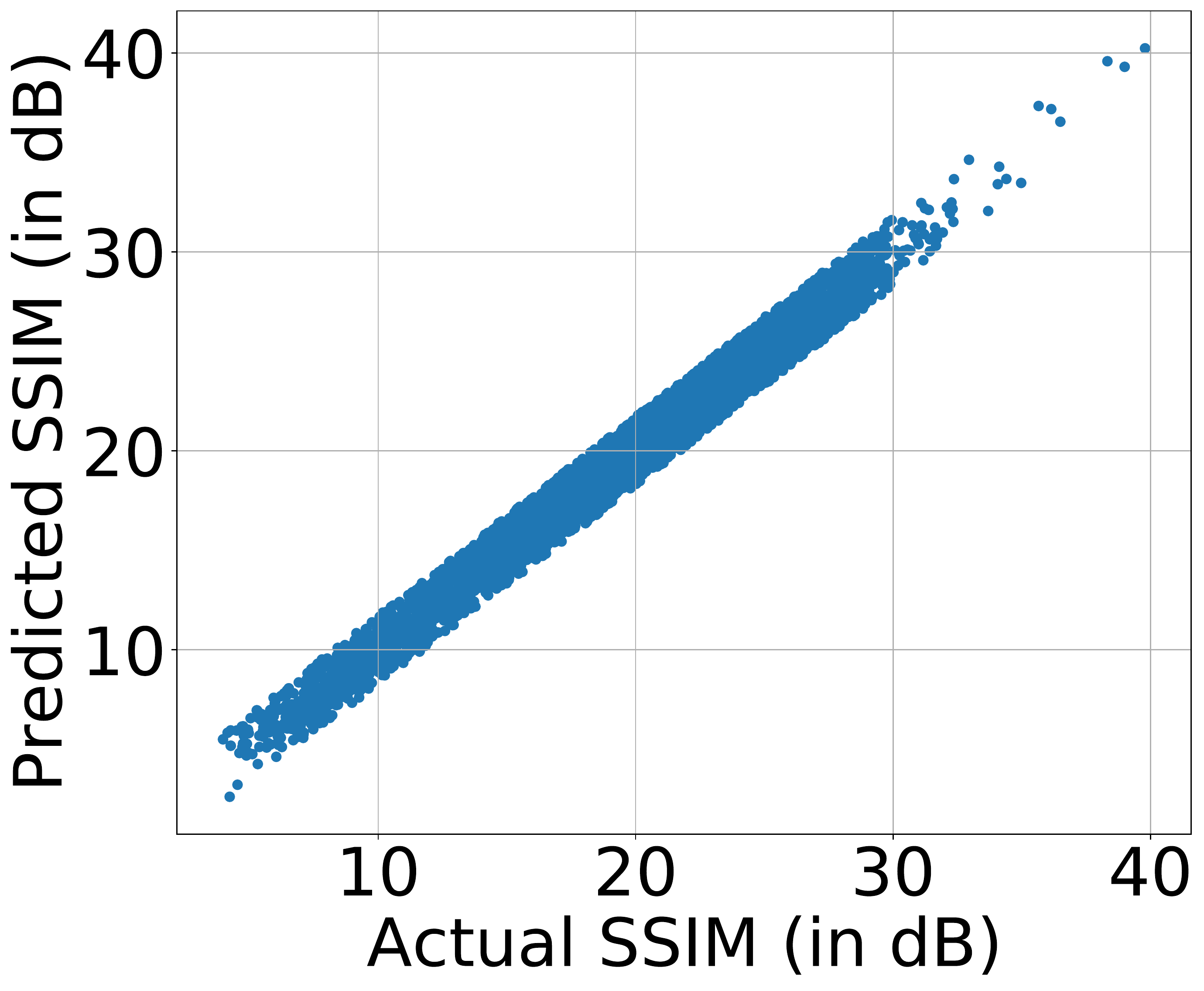}
    \caption{}
    \label{fig:ssim_m1_scatter}    
\end{subfigure}
\hfill
\begin{subfigure}{0.24\textwidth}
    \centering
    \includegraphics[width=\textwidth]{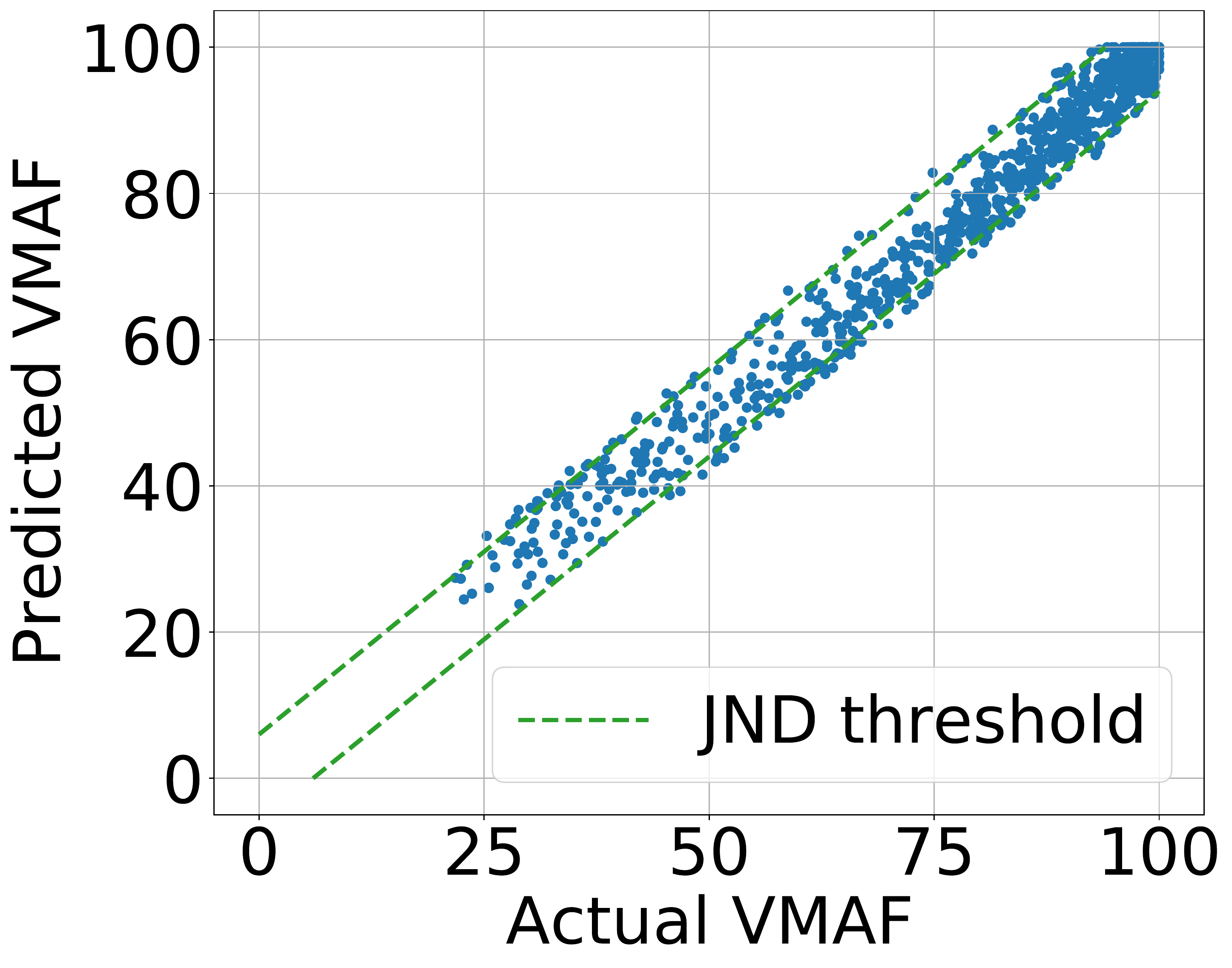}
    \caption{}
    \label{fig:vmaf_m1_scatter}    
\end{subfigure}
\vfill
\begin{subfigure}{0.24\textwidth}
    \centering
    \includegraphics[width=\textwidth]{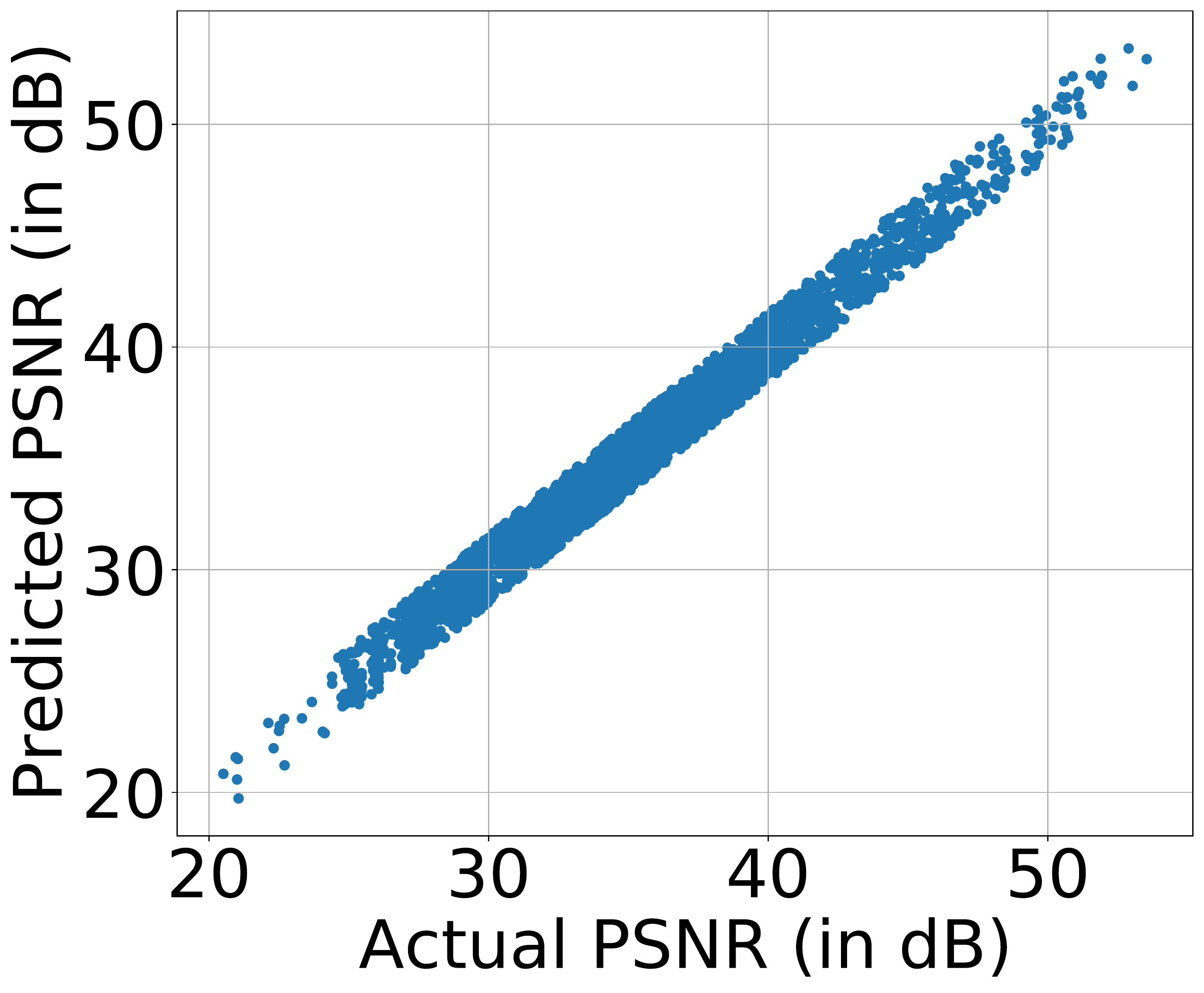}
    \caption{}
    \label{fig:psnr_m2_scatter}    
\end{subfigure}
\hfill
\begin{subfigure}{0.24\textwidth}
    \centering
    \includegraphics[width=\textwidth]{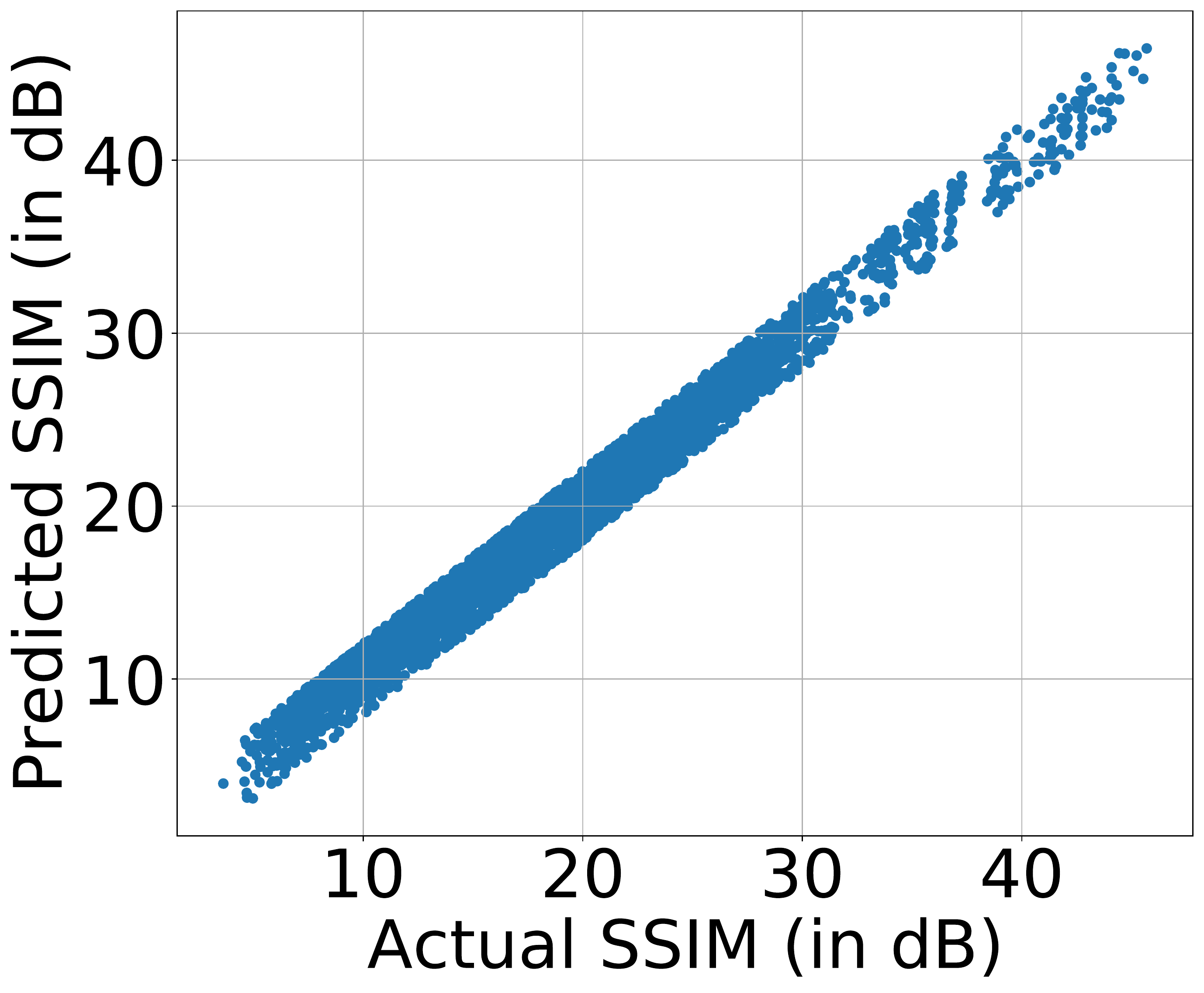}
    \caption{}
    \label{fig:ssim_m2_scatter}    
\end{subfigure}
\hfill
\begin{subfigure}{0.24\textwidth}
    \centering
    \includegraphics[width=\textwidth]{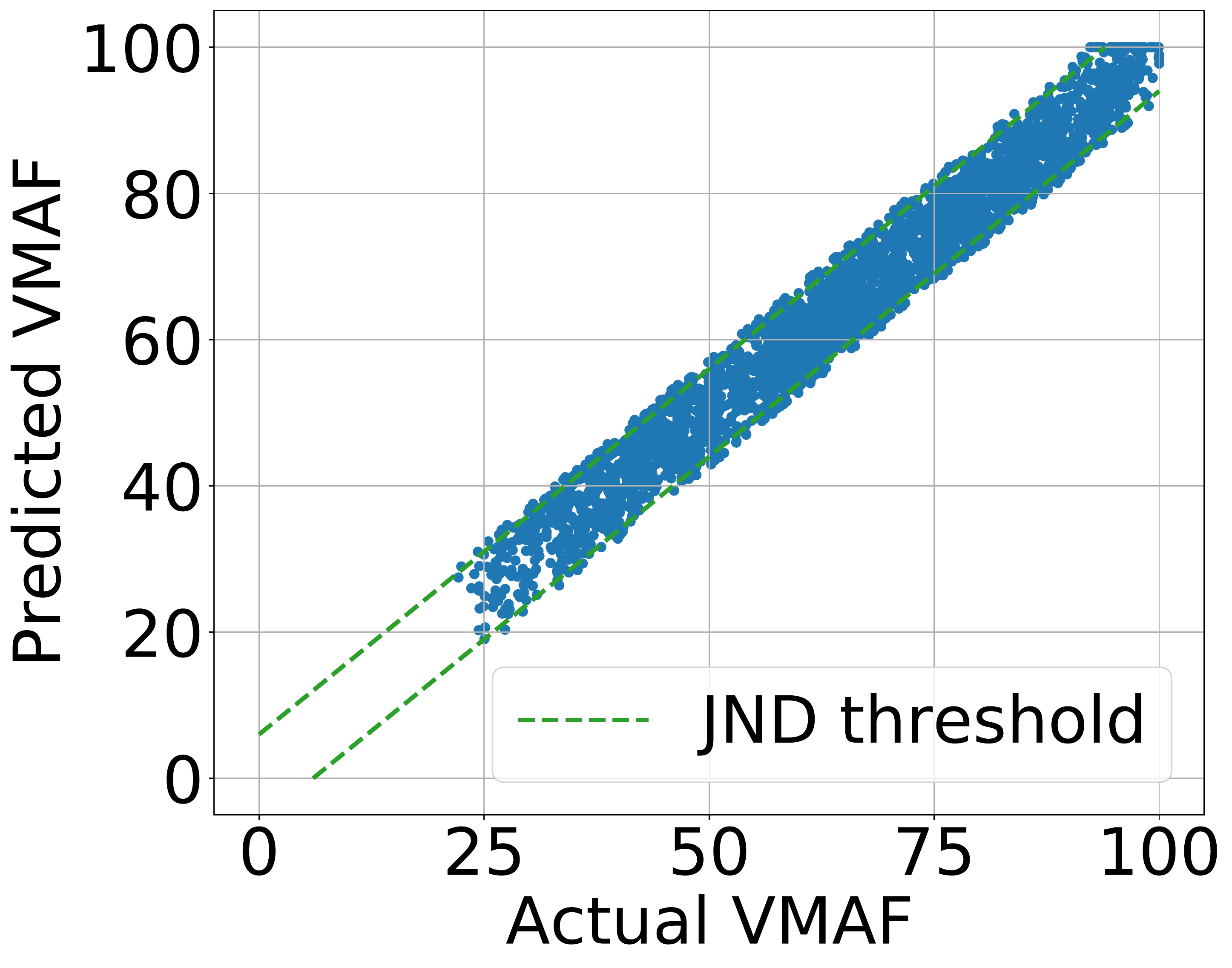}
    \caption{}
    \label{fig:vmaf_m2_scatter}    
\end{subfigure}
\caption{Scatterplots of the actual quality and predicted quality for \texttt{M=1} ((a) PSNR, (b) SSIM, and (c) VMAF, respectively) and \texttt{M=2} transcoding ((d) PSNR, (e) SSIM, and (f) VMAF, respectively).}
\label{fig:main_scatter_plot}
\end{figure*}
\begin{table*}[!t]
\caption{Prediction accuracy of \tvpm when \texttt{M=1} and \texttt{M=2}, respectively, for $\Tilde{b}_{1}$ representations considered in this paper encoded using x265 HEVC encoder.}
\centering
\resizebox{0.795\textwidth}{!}{
\begin{tabular}{l |c |c |c |c |c |c |c |c |c |c |c |c |c |c}
\multicolumn{3}{c|}{} & \multicolumn{4}{c|}{PSNR prediction} & \multicolumn{4}{c|}{SSIM prediction} & \multicolumn{4}{c}{VMAF prediction} \\
\specialrule{.12em}{.05em}{.05em}
\multicolumn{3}{c|}{} & \multicolumn{2}{c|}{\texttt{M=1}} & \multicolumn{2}{c|}{\texttt{M=2}} & \multicolumn{2}{c|}{\texttt{M=1}} & \multicolumn{2}{c|}{\texttt{M=2}} & \multicolumn{2}{c|}{\texttt{M=1}} & \multicolumn{2}{c}{\texttt{M=2}}\\
\specialrule{.12em}{.05em}{.05em}
\multicolumn{3}{c|}{$\Tilde{b}_{1}$} & $R^{2}$ & MAE & $R^{2}$ & MAE & $R^{2}$ & MAE & $R^{2}$ & MAE & $R^{2}$ & MAE & $R^{2}$ & MAE \\
\specialrule{.12em}{.05em}{.05em}
$b_{1}$ & 360p & 0.145 Mbps & 0.82 & 1.20 dB & - & -  & 0.89 & 1.08 dB & - & -  & 0.87 & 3.35 & - & - \\
$b_{2}$ & 432p & 0.300 Mbps & 0.83 & 1.19 dB & 0.84 & 1.37 dB & 0.89 & 1.14 dB & 0.87 & 1.34 dB & 0.87 & 3.51 & 0.76 & 3.38 \\
$b_{3}$ & 540p & 0.600 Mbps & 0.83 & 1.19 dB & 0.85 & 1.28 dB & 0.88 & 1.18 dB & 0.85 & 1.21 dB & 0.90 & 4.05 & 0.84 & 3.55 \\
$b_{4}$ & 540p & 0.900 Mbps & 0.83 & 1.19 dB & 0.83 & 1.22 dB & 0.86 & 1.17 dB & 0.86 & 1.11 dB & 0.90 & 3.83 & 0.89 & 3.53 \\
$b_{5}$ & 540p & 1.600 Mbps & 0.82 & 1.22 dB & 0.82 & 1.15 dB & 0.84 & 1.19 dB & 0.85 & 1.38 dB & 0.90 & 3.45 & 0.90 & 3.44 \\
$b_{6}$ & 720p & 2.400 Mbps & 0.83 & 1.26 dB & 0.83 & 1.28 dB & 0.82 & 1.18 dB & 0.83 & 1.57 dB & 0.88 & 2.88 & 0.91 & 3.45 \\
$b_{7}$ & 720p & 3.400 Mbps & 0.81 & 1.30 dB & 0.85 & 1.23 dB & 0.83 & 1.20 dB & 0.82 & 1.35 dB & 0.84 & 2.89 & 0.94 & 3.03 \\
$b_{8}$ & 1080p & 4.500 Mbps & 0.84 & 1.28 dB & 0.83 & 1.28 dB & 0.88 & 1.23 dB & 0.82 & 1.34 dB & 0.87 & 2.28 & 0.95 & 3.03 \\
$b_{9}$ & 1080p & 5.800 Mbps & 0.86 & 1.31 dB & 0.87 & 1.42 dB & 0.83 & 1.29 dB & 0.86 & 1.30 dB & 0.87 & 2.23 & 0.95 & 3.34 \\
$b_{10}$ & 1440p & 8.100 Mbps & 0.84 & 1.39 dB & 0.81 & 1.41 dB & 0.87 & 1.29 dB & 0.87 & 1.32 dB & 0.85 & 2.73 & 0.96 & 2.96 \\
$b_{11}$ & 2160p & 11.600 Mbps & 0.79 & 1.50 dB & 0.82 & 1.31 dB & 0.88 & 1.17 dB & 0.84 & 1.32 dB & 0.82 & 2.58 & 0.96 & 3.02 \\
$b_{12}$ & 2160p & 16.800 Mbps & 0.84 & 1.49 dB & 0.79 & 1.26 dB & 0.88 & 1.19 dB & 0.86 & 1.35 dB & 0.86 & 2.38 & 0.96 & 2.99 \\
\specialrule{.12em}{.05em}{.05em}
\multicolumn{3}{c|}{\textbf{Average}} & \textbf{0.83} & \textbf{1.31 dB} & \textbf{0.84} & \textbf{1.32 dB} & \textbf{0.85} & \textbf{1.19 dB} & \textbf{0.86} & \textbf{1.33 dB} & \textbf{0.87} & \textbf{3.01} & \textbf{0.91} & \textbf{3.25} \\
\specialrule{.12em}{.05em}{.05em}
\end{tabular}}
\label{tab:tvpm_res}
\end{table*}

\subsection{Evaluation Setup}
\label{sec:test_methodology}
In this paper, video sequences from JVET~\cite{jvet_video_ref}, MCML~\cite{mcml_video_ref}, SJTU~\cite{sjtu_video_ref}, Berlin~\cite{berlin_ref}, UVG~\cite{uvg_ref}, BVI~\cite{bvi-ref} datasets are used. The sequences are encoded at 30 fps using x265 v3.5$^{\ref{x265_ref}}$ with the \textit{ultrafast} preset using the \textit{Video Buffering Verifier} (VBV) rate control mode on a dual-processor server with Intel Xeon Gold 5218R (80 cores, frequency at 2.10 GHz). The segment length is set as four seconds. 80\% of the five hundred videos considered are used as the training dataset, and the remaining 20\% is used as the test dataset.  The bitrate representations considered in the experiments ($b_{j}~\forall~j \in [1, 12]$) used as the target bitrate of encoding in each transcoding stage ($\Tilde{b}_{i}~\forall~i \in [1, M]$) are specified in the Apple HLS authoring specifications$^{\ref{apple_hls_ref}}$. The $E$, $h$, and $L$ features are extracted using the VCA v2.0\footnote{\href{https://vca.itec.aau.at}{https://vca.itec.aau.at}, last access: Apr 02, 2023.} open-source video complexity analyzer~\cite{vca_ref} run in eight CPU threads, with $w$ (\cf Eq.~\ref{eq:block_energy}) as 32. $f_c$ is set as 15, \ie the video segment is divided into eight chunks (\texttt{T=8}).

Hyperparameter tuning is performed on the LSTM model to obtain the maximum prediction performance~\cite{lstm_survey_ref}. The number of LSTM cells is \rf{set} to 50, and the model is trained for 100 epochs with a learning rate of $10^{-3}$ with the Adam optimizer~\cite{adam_ref}. The loss function used to train \rf{the LSTM model} is the mean absolute error (MAE). The resulting quality and the predicted in terms of PSNR, SSIM, and VMAF~\cite{VMAF} are compared for each test sequence for \texttt{M=1} (single-stage) and \texttt{M=2} (two-stage) transcoding. Since the content is assumed to be displayed in the highest resolution (\ie 2160p), the transcoded content is scaled (bi-cubic) to 2160p resolution to determine the visual quality.

\subsection{Experimental Results}
\label{sec:exp_results}
In the first experiment, \tvpm's processing time (\ie $\tau_p$) is compared to the total transcoding latency $\tau_{T}$  (\cf Eq.\ref{eq:tot_ref_lat}) in \sota RR-VQA approaches. The average $\tau_{T}$ for \texttt{M=1} and \texttt{M=2} are observed as 1.92s and 3.78s, respectively. The average time taken for feature extraction ($\tau_{f}$ of a 4s segment is 0.323s. Furthermore, the average inference time of the LSTM model is 5 ms. Hence, the average processing time of \tvpm for a 4s segment is 0.328s. Thus, \tvpm has a significantly lower processing time than the \sota RR-VQA approaches. 

The second experiment assesses the correlation between the predicted to actual quality score for \texttt{M=1} and \texttt{M=2} transcoding. As illustrated in Figures~\ref{fig:psnr_m1_scatter}, \ref{fig:ssim_m1_scatter}, and \ref{fig:vmaf_m1_scatter} and Figures~\ref{fig:psnr_m2_scatter},~\ref{fig:ssim_m2_scatter}, and \ref{fig:vmaf_m2_scatter}, there is a strong correlation between the predicted to the actual PSNR, SSIM, and VMAF scores, respectively (\eg the average $R^{2}$ scores of VMAF prediction for single-stage and two-stage transcoding are 0.87 and 0.91, respectively). Furthermore, the prediction errors are less than the acceptable threshold of one JND (\ie six VMAF points, \rf{which shows \tvpm works with sufficient accuracy.}

In the final experiment, the prediction performance of \tvpm for the $\Tilde{b}_{1}$ representations considered in this paper is investigated using the Mean Absolute Error (MAE) for \texttt{M=1} and \texttt{M=2} transcoding. \rf{As shown in Table~\ref{tab:tvpm_res}}, the average MAE for VMAF prediction in \texttt{M=1} and \texttt{M=2} transcoding are 3.01 and 3.25, respectively. The results of \texttt{M=2} correspond to the average visual quality prediction accuracy of transcoding from $\Tilde{b}_{1}$ bitrate representation to the possible lower bitrate representations in the bitrate ladder. Please note that since $b_{1}$ is the lowest bitrate representation in the bitrate ladder, a scenario corresponding to $\Tilde{b}_{1} = b_{1}$ does not exist. The $R^{2}$ scores for \texttt{M=2} are observed to increase as $\Tilde{b}_{1}$ increases. This is because there is a higher amount of training data (transcoding to lower bitrate representations) as $\Tilde{b}_{1}$ increases. 

\section{Conclusions}
\label{sec:conclusion_future_dir}
This paper proposed \tvpm, an online transcoding quality prediction model for video streaming applications. The proposed \rf{LSTM-based} model uses DCT-energy-based features as \textit{reduced reference} to characterize the input video segment, which is used to predict the visual quality of an M-stage transcoding process. The performance of \tvpm is validated by the Apple HLS bitrate ladder encoding and transcoding using the x265 open-source HEVC encoder. On average, for single-stage transcoding, \tvpm predicts PSNR, SSIM, and VMAF with an MAE of 1.31 dB, 1.19 dB, and 3.01, respectively. Furthermore, PSNR, SSIM, and VMAF are predicted for two-stage transcoding with an average MAE of 1.32 dB, 1.33 dB, and 3.25, respectively.

In this paper, trans-sizing and trans-rating are considered as transcoding, \ie the encoder/codec used for the bitrate ladder representations is assumed to be the same. In the future, transcoding between bitrate ladder representations of various codecs shall be investigated. Another future direction is defining a decision-making component based on the proposed model in an end-to-end live streaming system.

\section{Acknowledgment}
The financial support of the Austrian Federal Ministry for Digital and Economic Affairs, the National Foundation for Research, Technology and Development, and the Christian Doppler Research Association is gratefully acknowledged. Christian Doppler Laboratory ATHENA: \small{\url{https://athena.itec.aau.at/}}.
\balance

\bibliographystyle{ACM-Reference-Format}
\bibliography{references.bib}
\balance
\end{document}